## Dynamical Equilibrium of Interacting Ant Societies

Boon Tiong Melvin Leok
MSC 609, Caltech, Pasadena, CA 91126-0609, USA
*mleok@cco.caltech.edu*
(818)395-1709

# Abstract

*The sustainable biodiversity associated with a specific ecological niche as a function of land area is analysed computationally by considering the interaction of ant societies over a collection of islands. A power law relationship between sustainable species and land area is observed. We will further consider the effect a perturbative inflow of ants has upon the model.*

# 1 Introduction

It has been observed qualitatively that the biodiversity of a landmass varies as a function of the area raised to a power less than unity. We will attempt to develop a numerical model of the interaction of ant societies which are constantly perturbed by the external inflow of new ants, and determine the sustainable number of species as a function of area in the dynamically stable equilibrium condition.

# 2 Model Representation

The conventions which will be utilized in the development of the numerical model of ant society interaction are described herein. They generally serve to constrain the scope and complexity of the model.

## 2.1 Phase and State Variables

The model will utilize phase variables to represent the independent parameters such as land area, and rate of inflow of new ants. State variables will represent the dynamic state of the simulation at a given time. These variables are described in the following section.

### 2.1.1 Discretization of time

This model will represent the dynamic state of ant populations using discrete time, thus the model will be formulated in the form of a set of coupled mappings.

### 2.1.2 Normalization of variables

Certain phase and state variables, such as the land area, population density will be normalized to the interval $[0, 1] \in \mathbb{R}$ as a method of simplifying the analysis of the model.

### 2.1.3 Representation of Populations

The state of the ant societies will be represented in matrix form, with the rows corresponding to the distinct ant species, and the columns corresponding to the island upon which they reside. As such the population state at a given time can be represented in the following matrix,

$$\mathbf{P}(t) = \begin{pmatrix} p_{11} & \cdots & p_{1m} \\ \vdots & \ddots & \vdots \\ p_{n1} & \cdots & p_{nm} \end{pmatrix} \tag{1}$$

where $p_{ij}$ = population of $i$-th ant species on the $j$-th island, $t \in \mathbb{N}$.

Associated with this population matrix, is the population density matrix, which is,

$$\mathbf{Pd}(t) = \begin{pmatrix} pd_{11} & \cdots & pd_{1m} \\ \vdots & \ddots & \vdots \\ pd_{n1} & \cdots & pd_{nm} \end{pmatrix} \tag{2}$$

where $pd_{ij}$ = population density of $i$-th ant species on the $j$-th island, $t \in \mathbb{N}$. The population density is related to the population by,

$$pd_{ij} = \frac{p_{ij}}{Area_j} \tag{3}$$

where $Area_j$ = maximum population for $j$-th island.

## 2.2 Initial conditions

The islands are assumed to be initially uninhabited, with the ant population on each island due to both self-propagation and the influx of new ants from the continent. In addition, the following conditions are assumed for the ant influx, and the self-interaction on each island.

### 2.2.1 Distribution of inflow population

It shall be assumed that the perturbative inflow of new ants from the continent will be randomly chosen from uniformly distributed pool of ant species. We shall further assume that the number of ants which are derived from immigration from the continent is proportional to the area of the island.

Due to the negligible size of the islands compared to that of the continent, the effect of backflow of ant populations to the continent shall be neglected.

#### 2.2.2 Genetic Stasis

We shall assume that the biodiversity is completely derived from the inflow of a variety of ant species from the continent and that the ant species do not adapt and deviate genetically from species obtained from the continent.

## 2.3 Population Interactions

The problem can be considered to be a generalization of the insect population model considered by May [1], in his paper which introduces the notion of the logistic mapping. The logistic map was originally used to represent the effect limited resources have upon the growth of insect populations.

The addition of interaction between competing ant populations can be modelled with the use of an inter-population attrition factor, which is described further below.

#### 2.3.1 Inter-island Interaction

The mode of inter-island interaction will be that of travel over water. Without loss of generality, the attrition of ant populations due to travel over water can be assumed to be 50% unit length$^{-1}$,

$$p(d) = p_0(0.5)^d \tag{4}$$

This relationship will serve to affect terms in the interaction matrix for ant populations residing on different islands.

The inter-island distances can be represented as a distance matrix as follows,

$$\mathbf{D} = \begin{pmatrix} d_{11} & \cdots & d_{1m} \\ \vdots & \ddots & \vdots \\ d_{m1} & \cdots & d_{mm} \end{pmatrix} \tag{5}$$

where $d_{ij}$ = distance between $i$-th and $j$-th island.

From the distance matrix, we can obtain the associated dissipation matrix, which determines the percentage of ants surviving the trip between the $i$-th and $j$-th island.

$$\begin{aligned} f_{ij} &= \frac{p(d_{ij})}{p_0} \\ &= (0.5)^{d_{ij}} \end{aligned} \tag{6}$$

#### 2.3.2 Inter-species Interaction

The ant societies will interact according to an interaction matrix. The intensity of the attrition will be dependent upon the population density of the island, such that an increase in the population density would increase both the possibility of confrontation between populations on the same island, as well as the possibility that a portion of the ant population ventures out in search of other island to colonize.

As each species potentially interacts with all other ant colonies both within the same island, and on different islands, we can define the following mapping,

$$p_{ij}(t+1) = f_{oj} \cdot N\left(\frac{Area_j}{n}, \frac{Area_j}{4n}\right) + \sum_{\substack{k=1..n \\ l=1..m}} a_{ijkl} p_{kl}(t) \tag{7}$$

where $f_{oj}$ corresponds to the dissipation due to travel from the continent to the $j$-th island. $N\left(\frac{Area_j}{n}, \frac{Area_j}{4n}\right)$ is a random variable from a normal distribution. The normally distributed random term corresponds to the contribution of immigration from the continent.

For the second term of the expression, $a_{ijkl}$ = interaction factor between $i$-th ant species on $j$-th island, and $k$-th ant species on $l$-th island.

#### 2.3.3 Strength of Interaction

We will be concerned with four possible sources of population change,

1. *internal population growth*
2. *attrition due to interaction with other species*
3. *population decrease due to emigration*
4. *population increase due to immigration, both from the continent and other islands*

The following functions serve to represent the intensity of population interactions. All of these functions will defined in terms of the total population density of the island, $x = \sum_{q=1}^{n} pd_{qj}$.

**Growth Function** The logistic mapping considers the interaction of a population with limiting environmental resources. We shall thus base the growth function on the same concept.

We shall represent the *total* population density at time $t$ on the $j$-th island as $T_j(t) = \sum_{q=1}^{n} pd_{qj}(t)$, and the next iterate of the *total* population density due to interaction only with the environment at time $t+1$ on the $j$-th island as $T_j(t+1) = \sum_{q=1}^{n} pd_{qj}(t+1)$.

Then by analogy to the logistic mapping, the following equation should be satisfied,

$$T_j(t+1) = \lambda \cdot T_j(t) \cdot [1 - T_j(t)] \tag{8}$$

Where $\lambda \in [0, 1]$.

We would like the *growth function* to satisfy the following,

$$p_{ij}(t+1) = p_{ij}(t) \cdot \frac{T_j(t+1)}{T_j(t)} = p_{ij}(t) \cdot G(x) \quad (9)$$

From which, we can derive the *growth function* as,

$$G(x) = \frac{T_j(t+1)}{T_j(t)} = \lambda \cdot [1 - T_j(t)] = \lambda \cdot (1 - x) \quad (10)$$

**Attrition Function** A is the *attrition function*, defined in this fashion,

$$A(x) = k_1 \cdot x^{\alpha} \quad (11)$$

**Emigration Function** E is the *emigration function*, which is related to the *attrition function* (11) by,

$$E(x) = \mu \cdot A(x) \quad (12)$$

where $\mu$ is a parameter within the range $[0, 1] \in \mathbb{R}$.

#### 2.3.4 Interaction Factors

There are generally four classes of inter-population interaction,

1. *self-interaction*
2. *interaction with different species on the same island*
3. *interaction with same species on different islands*
4. *interaction with different species on the same island*

The interaction factor for each of these cases will be considered in the following section.

**Self-Interaction** The interaction factor for self-interaction can be defined as the increase due to interaction population growth, less the decrease due to emigration. Thus,

$$a_{ijij} = G\left(\sum_{q=1}^{n} pd_{qj}\right) - E\left(\sum_{q=1}^{n} pd_{qj}\right) \quad (13)$$

**Interaction with different species on the same island** The effect interaction with different species on the same island have upon the population of a species in question, would be a decrease in population due to attrition.

$$a_{ijkj} = -A\left(\sum_{q=1}^{n} pd_{qj}\right) \quad (14)$$

**Interaction with same species on different islands** The increase due to immigration of the same species from different islands would be the portion of the immigrants multiplied by the dissipation due to travel.

$$a_{ijil} = f_{jl} \cdot E\left(\sum_{q=1}^{n} pd_{ql}\right) \quad (15)$$

**Interaction with different species on different islands** The attrition due to interaction with hostile immigrant species would be the attrition factor, multiplied by the portion of immigrants, multiplied by the dissipation due to travel.

$$a_{ijkl} = -f_{jl} \cdot E\left(\sum_{q=1}^{n} pd_{ql}\right) \cdot A\left(\sum_{q=1}^{n} pd_{qj}\right) \quad (16)$$

**Complete Formulation of the Interaction Factor** We can collate the interaction factors for the individual cases in this complete formulation,

$$a_{ijkl} = \begin{cases} G\left(\sum_{q=1}^{n} pd_{qj}\right) - E\left(\sum_{q=1}^{n} pd_{qj}\right) & i = k, j = l \\ f_{jl} \cdot E\left(\sum_{q=1}^{n} pd_{ql}\right) & i = k, j \neq l \\ -A\left(\sum_{q=1}^{n} pd_{qj}\right) & i \neq k, j = l \\ -f_{jl} \cdot E\left(\sum_{q=1}^{n} pd_{ql}\right) \cdot A\left(\sum_{q=1}^{n} pd_{qj}\right) & i \neq k, j \neq l \end{cases} \quad (17)$$

Thus, the above equation, in conjunction with the mapping (7), defines a mapping,

$$\mathbf{M} : \mathbf{P}(t) \rightarrow \mathbf{P}(t+1) \quad (18)$$

## 3 Implementation of the model

The formulation above describes the mathematical framework for the model of interacting ant populations. In order to implement the model, certain parameters and initial conditions need to be specified. These are described in the following section..

### 3.1 User-defined parameters

The following parameters need to be specified to uniquely define the simulation.

#### 3.1.1 Stability of Population Fluctuations

The stability of population fluctuations is influenced by the choice of the parameter $\lambda$. Recall that this parameter is analogous to the control parameter in the logistic mapping. As we're more concerned with inter-species interaction, we shall limit $\lambda$ to values in which the single population case results in a stable population, thus $\lambda$ is within the range $(1,3) \in \mathbb{R}$. For the purpose of the simulation, we shall let $\lambda = 2$.

#### 3.1.2 Competition between species

The competition between species results in attrition of the ant populations. $k_1$ corresponds to the percentage of the population of the island lost to conflict when the island is filled to capacity. $\alpha$ specifies the power law relation which the intensity of attrition increases as a function of population density. We shall be using $k_1 = 0.05$, $\alpha = 2.5$.

#### 3.1.3 Emigration

The emigration parameter $\mu$ determines the proportion of the population which are driven off the island due to population pressure. Since we can assume that the population outflow is relatively insubstantial, we shall take $\mu = 0.1$.

### 3.2 Initial Conditions

Initial conditions determine the physical conditions which the simulation evolves under. These include the distance from the continent, inter-island distances, and the areas of the individual islands.

#### 3.2.1 Distance from the continent

Distances from the continent should be sufficiently large that the influx of immigrant ants from the population do not overwhelm the inter-island dynamics of the system, we shall set the continental distance to a minimum of 5.

#### 3.2.2 Area of islands

The area of the individual islands will be randomly assigned to provide a representative sample of the phase space of simulations.

#### 3.2.3 Inter-island distances

The inter-island distances will be kept constant to reduce the effect of the placement of the randomly sized islands. Furthermore, to isolate the contribution of distances from the continent, all the inter-island distances will be set to 1.

## 4 Results from the simulation

The simulation was constrained to $n = 10$, $m = 5$. The first 100 iterations were discarded from the analysis, as they were assumed to exhibit transient as opposed to the characteristic dynamics of the system. The average number of species on each island over the next 100 turns was taken as the number of species which the island would support. We then fit the data to the power law relation to determine the scaling exponent, $\gamma$.

$$species_j \propto Area_j^{\gamma} \quad (19)$$

The scaling exponent is obtained as the gradient of a log-log plot of $species_j$ versus $area_j$. This was repeated over 20 initial area configurations, and the mean and standard deviation of the scaling exponents were obtained.

### 4.1 Scaling exponent as a function of distance

The scaling exponent for the number of species changes as a function of distance from the continent. The tabulated results are as follows.

| distance | $\overline{\gamma}$ | $\sigma$ |
|---|---|---|
| 5.00 | 0.3656 | 0.0732 |
| 5.25 | 0.3162 | 0.0921 |
| 5.50 | 0.2791 | 0.0839 |
| 5.75 | 0.2383 | 0.0862 |
| 6.00 | 0.2001 | 0.0570 |
| 6.50 | 0.1718 | 0.0532 |
| 7.00 | 0.1527 | 0.0812 |
| 8.00 | 0.1293 | 0.1109 |
| 10.00 | 0.1210 | 0.0869 |
| 14.00 | 0.1305 | 0.1134 |
| 18.00 | 0.1190 | 0.0470 |

The results are illustrated graphically in the plot below.

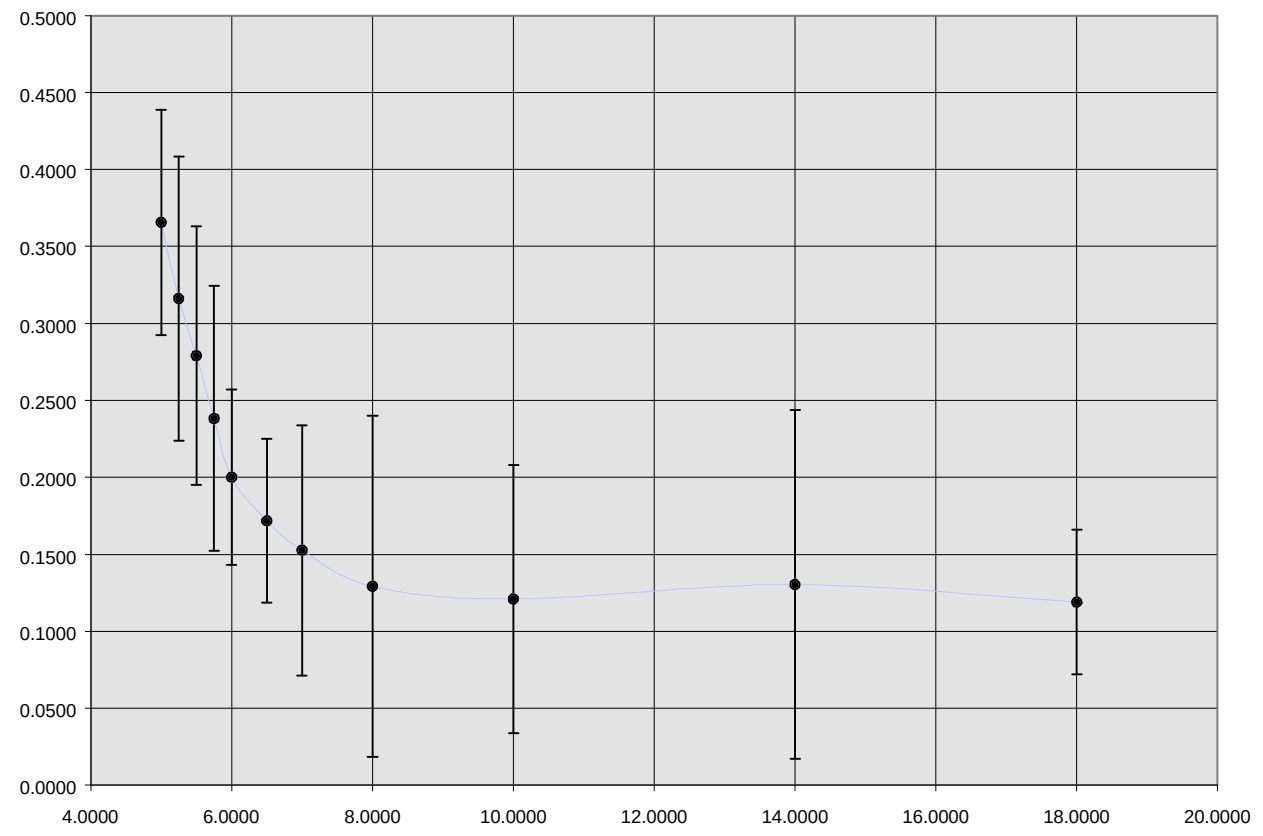

Scaling exponent *versus* distance

Note that there are two modes of saturation in the simulation. In the case of extremely low values of distance, the inflow of new ants is so substantial as to cause the number of species on each island to saturate at the maximum number of species available. In the extreme of high values of distance, the external perturbation is so small that the islands can be considered to be independent of the influence of the continent once the dynamics settle onto the attractor of the phase space.

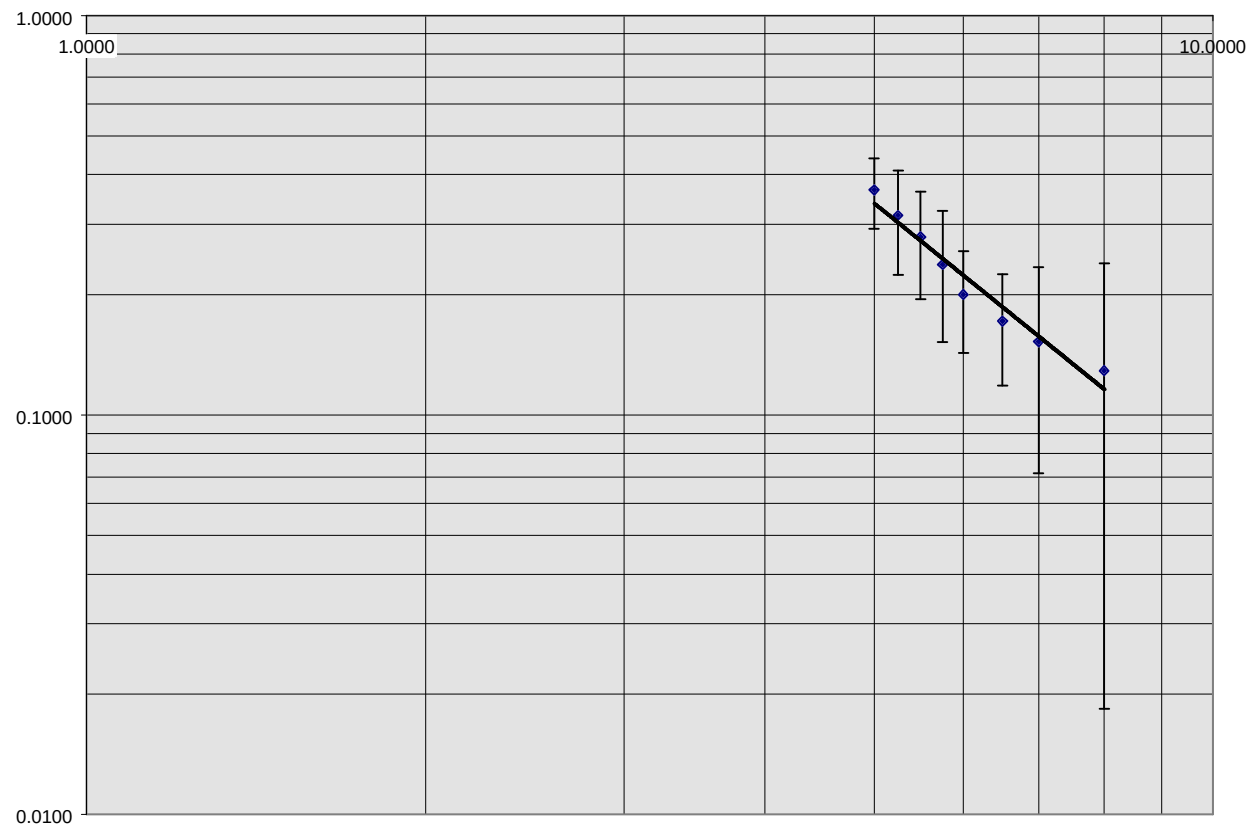

*log-log* plot of scaling exponent *versus* distance

The power law regression for the 8 data points with distances between 5 and 8 gives the following relationship for the scaling exponent and distance,

$$\gamma \propto distance^{-0.0766} \tag{20}$$

The $r^2 = 0.8442$ does suggest that the power law relation between the scaling exponent and distance from the continent is a reasonable approximation.

# 5 Conclusion

The results of our simulation tends to indicate that the dynamical equilibrium of interacting ant societies follows a power law relation between the number of species and the area of the island. The scaling exponent is less than unity, and varies according to the power law, $\gamma \propto distance^{-0.0766}$, for moderate values of distance.

The exact form of the power law relation between the scaling exponents and the distance is likely to vary depending on the user parameters specified for the simulation. However, the quality of the power law regression does suggest that for dynamics with qualitatively similar behaviour, an analogous power law relationship is likely to arise for distances which do not result in saturation.